\documentclass[journal,onecolumn,12pt]{IEEEtran}
\usepackage{cite}

\ifCLASSINFOpdf
\else
\fi
\usepackage{color}
\usepackage{placeins}
\usepackage{float}
\usepackage{tabularx,colortbl}
\usepackage{cite}
\usepackage{amsmath,amssymb,amsfonts}
\usepackage{algorithmic}
\usepackage{graphicx,epstopdf}
\usepackage{subfigure}
\usepackage{textcomp}
\usepackage{xcolor}
\usepackage{setspace}

\usepackage{amsmath}
\begin{document}
%
\title{Compact Design of Dual-Band Circular Polarized Microstrip Antenna with Single Feed}
%
%
%

\author{Sandip~Ghosal,
        Raed~M.~Shubair}
\maketitle

\begin{abstract}
A novel and compact dual band dual sense circularly polarized microstrip patch antenna with single coaxial feed has been reported in the present work. The key idea of generating dual band circular polarisation (CP) is the integration of a square patch with corner truncation and a smaller concentric circular patch with double slits. The first resonance is provided by the larger patch whose corner truncation generates two orthogonal modes. The inner patch controls the higher-order resonance with the CP contributed by two narrow slits.  The higher order resonating frequency can be monitored by controlling the dimensions of the circle and the slits. The antenna provides the CP in two orthogonal planes with two different sense of polarisation. The lower order CP is of left-handed orientation, whereas the higher order shows right-handed polarization. The cross-polarization level is also found to be very low.  
\end{abstract}


%
\IEEEpeerreviewmaketitle

\pagebreak
\linespread{1.25}
\section{Introduction}
\linespread{1.25}
%
%
%
%
\IEEEPARstart{I}{n} the present era of modern technology, antennas have become an integrated part of daily life applications such for medical purposes \cite{che2008propagation,omar2016uwb,shubair2015vivo,alhajri2015hybrid,al2006direction,al2003performance,al2003investigation,khan2017ultra,shubair2005robust,elayan2017wireless,shah2016posture,yang2018freezing,al2005direction,yang2018detection,shubair2015novel,khan2018frequency,nwalozie2013simple,shah2019cognitive} and industrial utility \cite{shah2017buried,khan2016compact,omar2016uwb}. Therefore, significant research has been going on to analyse and improve various antenna parameters. In this regard, So there exists different design parameters like -two frequencies of resonance, two orthogonal circular polarization vectors, input impedance at single feed location to match the antenna at both the resonating frequencies and the cross-polarization level. The frequency of resonance can be determined using the existing literature of  \cite{harrington1971theory,sandip_temc1}. Once the resonating frequencies are determined, corresponding input impedance can be computed the network formulation of \cite{ghosal2013analysis}. The polarisation of the field radiated from the antenna depends on the current induced on the antenna structure \cite{khan2016pattern,ibrahim2017compact}. Thereby, the radiated field characteristics can be controlled through systematic perturbation of the antenna geometry. Loading the radiating element with some narrow slot or slit is one popular technique of shape perturbation. In this, some recent studies have been reported on the slot loading effect on the antenna current polarization in \cite{ghosal2013analysis,mtechthesis,ghosal2018analysis,hussein2016novel,ghosal2018characteristic}.
In general, dual band circularly polarised antennas are found to be a popular choice for WLAN, Wi-Fi, and global positioning satellite applications \cite{khancircularly}. In the earlier stage, aperture coupled stacked microstrip patches were used to achieve the dual band CP operation in \cite{pozar1997dual,lau2005wide}. Such multi-layer patch topologies consume a higher amount of real estate along with a relatively complicated feeding network needs. Henceforth, the attention has been shifted to the bottom side of the antennas where either high impedance surface added below the ground \cite{cai2015dual} or the ground itself is loaded with variously shaped slots to generate circular polarisation in two frequency bands \cite{bao2007dual,ko2013dual}. Various metasurface and other metamaterial inspired structures are also being employed to design dual band CP antennas as found in \cite{li2015novel,chen2018dual}.
In parallel with the patch antennas, different monopole antennas were also studied for this purpose. In \cite{lu2014planar,komulainen2007frequency} planar dual-band 
monopole antenna was proposed where an inverted
L-shaped strip-sleeve was shorted to the ground plane to achieve circular polarisation in two different resonating frequencies. Alternately, a combination of  inverted C-shaped (ICS) and tilted I-shaped (TIS) structures were used to constitute a tilted-D-shaped monopole antenna which provides wide dual-band dual-sense circular polarization. Two annular rings were above a relatively larger height from the ground plane in \cite{mener2016dual} to obtain dual band dual CP operation in the K-band. Recently, some alternate techniques are also reported using multiple stacked layers or additional EBG layer or multiple elements in \cite{zeb2014high,yang2019compact,zhang2015dual} which have the main limitation of larger design volume. The presence of multiple layers or multiple elements lead to unwanted amount of coupling which affects the antenna's polarization behaviour \cite{mtechthesis,ghosh2016mutual,ghosal2019vtc,ghosal2019aps}.\\
Previously in \cite{mtechthesis}, the design of circularly polarised microstrip antennas has been extensively studied with various feeding techniques and patch geometries. It was found that a symmetry in the radiator's topology leads to the generation of two orthogonal degenerate modes. By varying the physical dimensional parameters , the frequency of resonance and sense of polarization can be controlled. As an extension of the existing theory \cite{mtechthesis}, the present work reports a dual band circularly polarised antenna with single coaxial probe feeding. The simple and miniaturized design makes it suitable for compact applications. The design and the results are discussed below.

\section{Results and Discussions}
\begin{figure}[H]%
\centering
\subfigure[Schematic]{%
\label{fig1a}%
\centering
\includegraphics[width=0.45\columnwidth,clip]{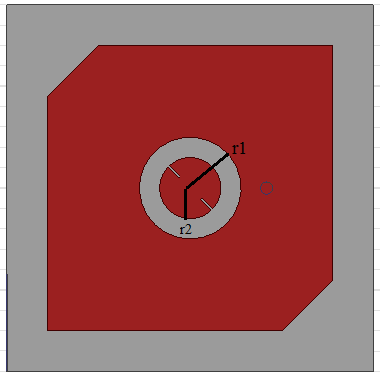}}
\hfill
\\
\subfigure[$S_{11}$ vs. frequency]{%
\label{fig1b}%
\centering
\includegraphics[width=0.45\columnwidth,clip]{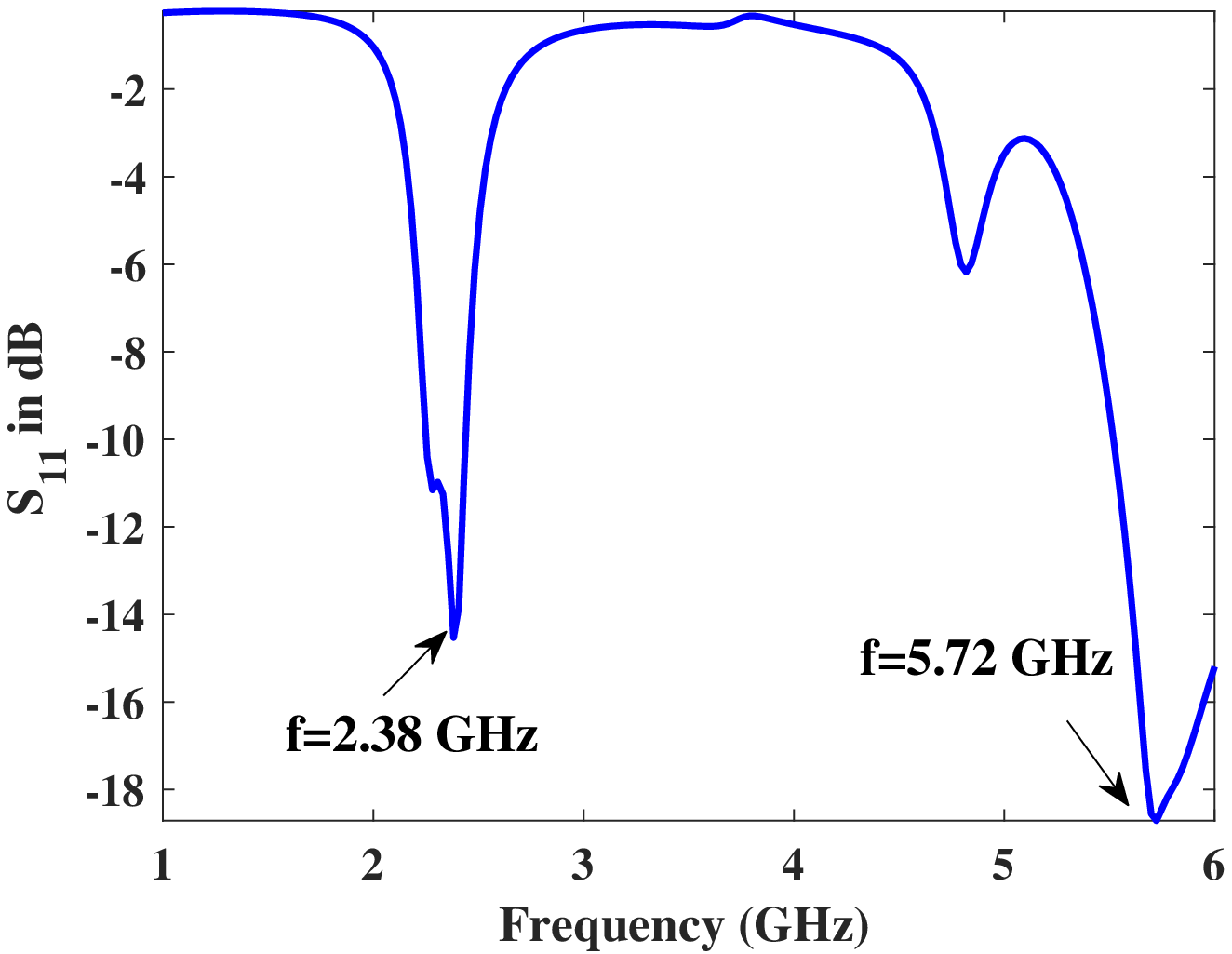}}
\caption{Top view of the proposed antenna and its frequency variation of $S_{11}$ parameter.}
\end{figure}

A common measure of circular polarisation is the axial ratio which is defined by the amplitude of the field strength in two orthogonal directions. The ideal value of the axial ratio for circular polarisation will be 1 in absolute magnitude or 0 dB in logarithmic scale. In general, antennas with an axial ratio of below 3 dB are termed as circularly polarised. To design a dual band CP antenna, a square-shaped microstrip antenna is designed initially with the side lengths of the patch as $28~\text{mm} \times 28~\text{mm}$. The dimension of the ground plane of the antenna is $36~\text{mm} \times 36~\text{mm}$. The height of the FR4 substrate ($\epsilon_r=4.4$) is 1.6 mm. The antenna is fed using coaxial probe feed. The square patch antenna shows linear polarization inherently. Then, two opposite corners of the antenna are truncated by 5 mm on both sides as shown in Fig. \ref{fig1a}. 

\begin{figure}[H]%
\centering
\subfigure[AR vs. frequency at4 $\phi=90^{\circ}$ plane.]{%
\label{fig2a}%
\centering
\includegraphics[width=0.45\columnwidth,clip]{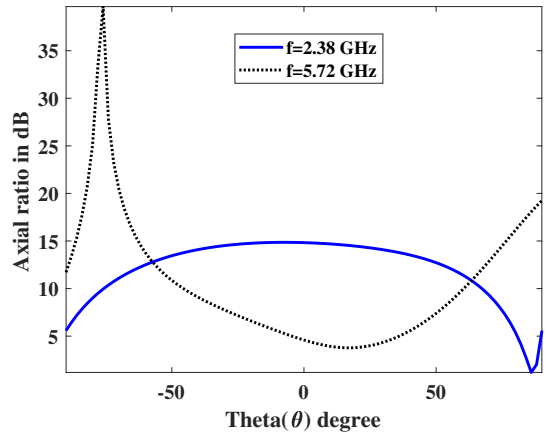}}
\hfill
\\
\subfigure[AR vs. frequency at4 $\phi=0^{\circ}$ plane.]{%
\label{fig2b}%
\centering
\includegraphics[width=0.45\columnwidth,clip]{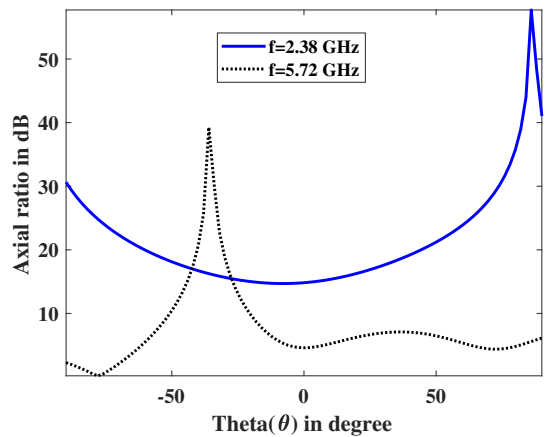}}
\caption{Comparison of Axial ration versus frequency variation in two orthogonal planes.}
\end{figure}

The change in the side length leads to a phase change of 90 degree generating two orthogonal modes. The corner truncation provides the lower order resonance with circular polarization. With no annular ring, the truncated corner patch antenna shows circular polarization at 2.38 GHz. For the sake of dual band operation, two narrow slits are introduced on the inner circular patch at -45 degree angle. This leads to the formation of a higher order resonating mode. By suitably varying the length and width of the narrow slits, the higher order circular polarisation band is generated. As the annular ring is loaded on the patch surface with two narrow slits, dual band resonance, as well as dual band CP feature, is generated. With the dimensions of  $r_1$ and  $r_2$ as 5 mm and 3 mm, the $S_{11}$ parameter is reported in Fig. \ref{fig1b} which shows two resonances at $f=2.38$ GHz and $5.72$ GHz. The variation of the axial ratio in two orthogonal planes is shown in Figs. \ref{fig2a} and \ref{fig2b}. 

\begin{figure}[H]%
\centering
\subfigure[$\phi=90^{\circ}, \theta=86^{\circ}$ direction]{%
\label{fig3a}%
\centering
\includegraphics[width=0.5\columnwidth]{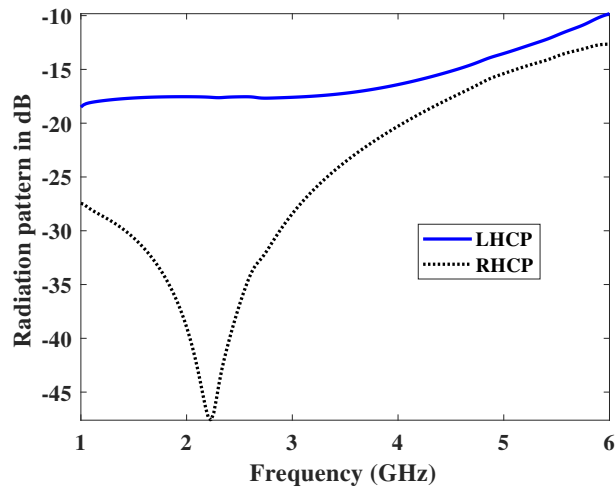}}
\hfill
\subfigure[$\phi=0^{\circ}, \theta=-78^{\circ}$]{%
\label{fig3b}%
\centering
\includegraphics[width=0.5\columnwidth]{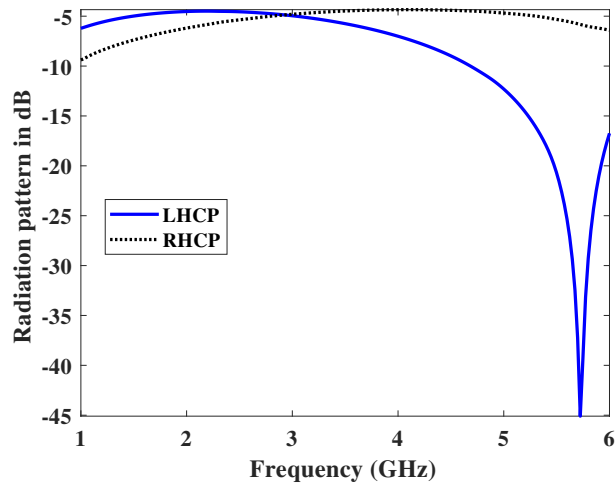}}
\caption{Comparison of the radiation patterns.}
\end{figure}

\begin{figure}[H]%
\centering
\subfigure[$2.38$ GHz ]{%
\label{fig4a}%
\centering
\includegraphics[width=0.5\columnwidth]{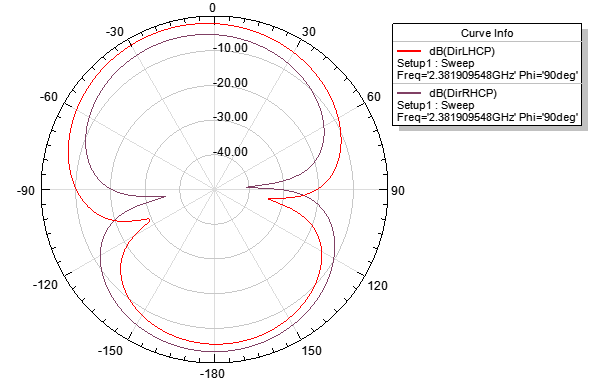}}
\hfill
\subfigure[$5.72$ GHz ]{%
\label{fig4b}%
\centering
\includegraphics[width=0.5\columnwidth]{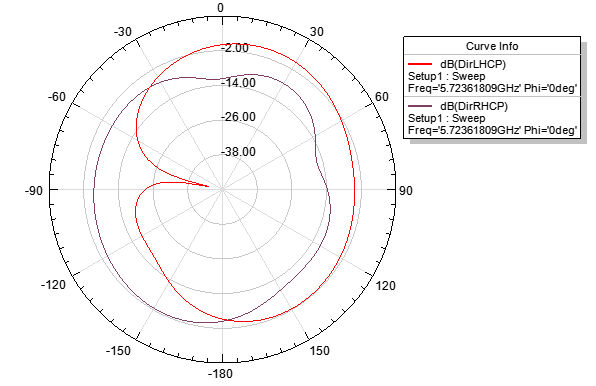}}
\caption{Comparison of the directivity patterns at two orthogonal planes. }
\end{figure}

It can be seen from Fig. \ref{fig2a} that at the lower order resonating frequency of 2.38 GHz, the circular polarisation arises at $\theta=86^{\circ}$ in the $\phi=90^{\circ}$ plane with the corresponding AR values of 1.18 dB. Alternately, at the higher order resonating frequency of 5.72 GHz, the circular polarisation is noted at  $\theta=-78^{\circ}$ in the $\phi=0^{\circ}$ plane in Fig. \ref{fig2b} with the AR value of 0.19 dB. To determine the sense of the polarisation, the radiation patterns are shown in Figs. \ref{fig3a} and \ref{fig3b}. The lower order resonance at $f=2.38$ GHz provides left-handed circular polarisation (LHCP) for both values of $\theta$ in Fig. \ref{fig3a}. The higher order resonance at $f=5.72$ GHz seems to generate right-handed circular polarisation (RHCP) in Fig. \ref{fig3b}.
It can be seen in Figs. \ref{fig3a} and \ref{fig3b} that the cross-polarisation component is quite low compared to its corresponding co-polarisation. Broadside peak gain values at two resonating frequencies $2.38$ GHz and $5.72$ GHz are -2.72 dB and 0.45 dB, respectively. The broadside patterns at $2.38$ GHz and  $5.72$ GHz are shown in Figs. \ref{fig4a} and \ref{fig4b}. It is worth of mentioning here that each polarisation orientation can be changed by reversing the corner truncation or the slit loading. All the full-wave simulations of the antennas have been carried using the FEM-based simulator HFSS \cite{hfss}.

\pagebreak

\section{Conclusion}
In the reported design, circular polarization was achieved at two bands with a single probe feeding. The lower band was generated due to the corner truncation of the square patch. The inner circular patch with two slits provided the higher band CP. The dual bands of CP were shown in two orthogonal planes. The lower order resonance was LHCP and the higher order one was RHCP. The antenna can find application in the polarisation diversity scheme of wireless communication.

\ifCLASSOPTIONcaptionsoff
  \newpage
\fi

\bibliographystyle{IEEEtran}
\bibliography{main}

\end{document}